\documentclass{aastex62}

\submitjournal{ApJ}

\shorttitle{Formation of 2A 1822$-$371}
\shortauthors{}

\begin{document}

\title{ON THE RAPID ORBITAL EXPANSION IN THE COMPACT LOW-MASS X-RAY BINARY 2A 1822$-$371}

\author{Ze-Pei Xing$^{1}$ and Xiang-Dong Li}
\affil{Department of Astronomy, Nanjing University, Nanjing 210023, China;
lixd@nju.edu.cn}

\affil{Key Laboratory of Modern Astronomy and Astrophysics, Nanjing University,
Ministry of Education, Nanjing 210023, China}

\begin{abstract}
The neutron star low-mass X-ray binary 2A 1822$-$371 has an orbital period of 5.57 hr. Mass transfer in such short-period binaries is thought to be driven by magnetic braking with orbital shrinking. However, 2A 1822$-$371 shows a very rapid orbital expansion, implying that mass transfer occurs rapidly in this system. The accretion rate of the neutron star is observationally estimated to be higher than the Eddington limit, which is also hard to be explained by the standard magnetic braking mechanism. In this work, we construct a model to account for the peculiar properties of 2A 1822$-$371. We assume that the donor star possesses a relatively strong magnetic field, which is coupled with the stellar winds excited by the X-ray radiation from the neutron star.  This would generate efficient angular momentum loss, leading to a high mass transfer rate and hence orbital expansion. We provide possible evolutionary tracks of 2A 1822$-$371 and study how the input parameters affect the results. The observational implications of the irradiation-driven mass loss are also briefly discussed  in the context of evolution of low-mass X-ray binaries and millisecond pulsars.

\end{abstract}

\keywords{pulsars: general - stars: evolution - stars: neutron - X-rays: binaries}


\section{Introduction} \label{sec:intro}

2A 1822$-$371 is a short-orbital period ($P_{\rm orb}=5.57$ hr) low-mass X-ray binary (LMXB) \citep{1980ApJ...242L.109M,1981ApJ...247..994W}. It hosts a neutron star (NS) with a spin period of $\sim 0.59$ s \citep{J01}, which is spinning up at a rate of $\sim -2.5\times 10^{-12}$ s\,s$^{-1}$ \citep{2010MNRAS.409..755,2015A&A...577A..63I}. Its partial eclipses imply that it is an accretion disk corona (ADC) source \citep{1982ApJ...257..318W} and the binary is viewed at a high inclination angle of 81\degr-85\degr \citep{2001MNRAS.320..249H,2011ApJ...729..102J}. Assuming a distance of 2.5 kpc\footnote{More recent {\em Gaia} observations indicate a distance of $8(\pm 4)$ kpc \citep{T18}.}, \citet{1982ApJ...262..253M} estimated an unabsorbed X-ray luminosity $L_{\rm X}\simeq 1\times 10^{36}$ erg s$^{-1}$, but the mean ratio of the X-ray over optical luminosity $L_{\rm X}/L_{\rm opt}$ for 2A 1822$-$371 is about 20. This is significant smaller than the typical value ($\sim 1000$) for LMXBs, suggesting that its intrinsic luminosity is likely higher than its Eddington limit \citep{2010ApJ...709..251B,2010A&A...515A..44B}. However, it is hard to account for such a high luminosity in compact LMXBs with the standard evolution theory (see below).

By studying the ``K-correction" for irradiated emission lines, \citet{2005ApJ...635..502M} obtained the masses of the NS and the donor star to be $1.61\,M_\sun\leq M_{\rm NS}\leq 2.32\,M_\sun$ and $0.44\,M_\sun\leq M_{\rm d}\leq 0.56\,M_\sun$, respectively. Based on the spin evolution of the NS, \citet{2015A&A...577A..63I} derived the NS mass to be  $1.69(\pm0.13)\,M_{\sun}$ by use of the magnetized accretion disk model of \citet{1979ApJ...234..296G}, which yields the donor mass to be $0.46(\pm0.02)\,M_{\sun}$.  \citet{2014PASJ...66...35S} reported a cyclotron resonance scattering feature (CRSF) at $33(\pm2)$ keV in the X-ray spectrum from a {\em Suzaku} observation of 2A 1822$-$371, suggesting that the surface magnetic field of the NS is $\sim 2.8(\pm0.2)\times 10^{12}\,\rm G$. \citet{2015A&A...577A..63I} analyzed the X-ray observational data of 2A 1822$-$371 made with {\em XMM-Newton}, {\em Chandra}, {\em Suzaku} and {\em INTEGRAL}
satellites. They did not detect any feature at $\sim 33$ keV but instead suggested a possible absorption feature near 0.7 keV, indicative a magnetic field strength of $8.8(\pm 0.3) \times 10^{10}\,\rm G$ for the NS. Nevertheless, the relatively long spin period and strong magnetic field of 2A 1822$-$371 compared with those of millisecond pulsars imply that it could be at the beginning stage of the recycling process.

Another remarkable feature of 2A 1822$-$371 is its orbital evolution.
Its orbital period was accurately measured through optical \citep{1979IAUC.3406....1S}, X-ray \citep{1981ApJ...247..994W}, infrared \citep{1982ApJ...255..603M} and ultraviolet \citep{1982ApJ...262..253M} observations.  \citet{1990MNRAS.244P..39H} estimated for the first time an orbital period derivative of $2.19(\pm0.58)\times10^{-10} $ s\,s$^{-1}$. \citet{2010ApJ...709..251B} studied the optical and UV data of 2A 1822$-$371 and obtained an orbital period derivative of $2.1(\pm0.2)\times10^{-10} $ s\,s$^{-1}$. By analyzing the X-ray data for a long time span, \citet{2010A&A...515A..44B}, \citet{2011A&A...534A..85I}, and \citet{2016ApJ...831...29C} derived an orbital period derivative of  $\sim 1.5\times10^{-10}$ s\,s$^{-1}$.  These values imply a very fast orbital expansion, anomalous for compact LMXBs in which mass transfer is driven by angular momentum loss via magnetic braking and gravitational radiation. The high positive orbital period derivative may indicate a high mass transfer rate $\geq 10^{-8}\,M_\sun$ yr$^{-1}$ \citep{2005ApJ...635..502M,2017MNRAS.468..824B}. However, the low mass of the companion star means that it is difficult to sustain such a mass transfer rate with conventional binary evolution theory.

A possible scenario to account for the high orbital period derivative is irradiation-driven mass transfer \citep{1993ApJ...410..281T,2016ApJ...831...29C}. One possible consequence of X-ray irradiation in LMXBs is mass loss from the donor star at a rate much higher than the traditional wind loss rate. In this case, if the donor star is a low-mass main-sequence star and possesses a sufficiently strong magnetic field, magnetic braking can efficiently extract orbital angular momentum and maintain rapid mass transfer. For example,
\citet{2016MNRAS.456..263P} modified the magnetic braking law for low-mass stars with strong winds to solve the high mass transfer rate problem in Sco X$-$1. \citet{2017A&A...606A..60C} also investigated the evolution Sco X$-$1, utilizing the anomalous magnetic braking scenario for Ap/Bp stars with strong magnetic fields proposed by \citet{2006MNRAS.366.1415}. Another possible effect of irradiation is thermal relaxation of the companion star.
The X-ray radiation energy from the NS could be deposited in the outer layer of the companion star. For low-mass main-sequence stars which have a convective envelope,  irradiation can block energy transport through the illuminated outer layers. The influence of irradiation on mass transfer in LMXBs was investigated by several authors \citep{P91,1997A&AS..123..273H,R95,K96}. \citet{2004A&A...423..281B} numerically calculated the long-term evolution of compact binaries with irradiation feedback and confirmed that compact LMXBs possibly undergo irradiation-driven mass transfer cycles. Recently, \citet{2018MNRAS.479..817T} simulated the evolutionary paths of SAX J1808.4$-$3658,  taking into account the effects of illumination of the donor by both the X-ray emission and the spin-down luminosity of the pulsar.

In this paper we examine the possible effects of irradiation on the evolution of 2A 1822$-$371 and attempt to explain its peculiar properties. In section 2, we describe the model for irradiation-driven mass transfer and orbital evolution. Section 3 presents the results of numerical calculation and the possible influence of input parameters. We discuss the observational implications for the LMXB evolution in section 4, and conclude in section 5.

\section{Wind-driven mass transfer} \label{sec:style}

We assume that the binary system contains a NS of mass $M_{{\rm NS}}$ and a donor star of mass $M_{{\rm d}}$ with Solar chemical compositions. We use the stellar evolution code Modules for Experiments in Stellar Astrophysics  \citep[MESA, ver. 10000;][]{2011ApJS..192....3P,2013ApJS..208....4P,2015ApJS..220...15P} to simulate the binary evolution. The effective Roche-lobe radius of the donor star is calculated by using the formula of \citet{1983ApJ...268..368E}:
\begin{equation}
\frac{R_{{\rm L}}}{a}=\frac{0.49q^{2/3}}{0.6q^{2/3}+\ln{(1+q^{1/3})}},
\end{equation}
where $q=M_{{\rm d}}/M_{{\rm NS}}$ is the mass ratio and $a$ is the orbital separation.

The mass transfer rate ($\dot M_{{\rm tr}}$) via Roche-lobe overflow (RLOF) is evaluated with the \citet{1988A&A...202...93R} scheme. We assume that a fraction $\beta$ of the transferred material, which carries the specific angular momentum of the NS, is lost from the NS.  Then the NS accretion rate can be written as
\begin{equation}
\dot M_{{\rm NS}} = (1-\beta)|\dot M_{{\rm tr}}|.
\end{equation}
We adopt $\beta=0.5$; meanwhile, we assume that $\dot M_{{\rm NS}}$ is limited by the Eddington accretion rate $\dot M_{{\rm Edd}}(\simeq 2 \times10^{-8}\,M_{\sun}$ yr$^{-1}$ for a $1.4\,M_{\sun}$ NS), so $\beta=0.5$ when $|\dot M_{{\rm tr}}|\leq 2\dot{M}_{{\rm Edd}}$, and $1 - \dot M_{{\rm Edd}}/|\dot M_{{\rm tr}}|$ when $|\dot M_{{\rm tr}}|> 2\dot{M}_{{\rm Edd}}$.
The accretion X-ray luminosity of the NS is
\begin{equation}
L_{{\rm x}}=\frac{GM_{{\rm NS}}\dot M_{{\rm NS}}}{R_{{\rm NS}}},
\end{equation}
where $R_{\rm NS}$ is the NS radius.
We further assume that X-ray radiation would drive a wind from the surface of the donor star \citep{1989ApJ...343..292R,1993ApJ...410..281T}. Assuming that a fraction $f$ of the accretion luminosity that the donor star receives is converted into the kinetic energy of the wind to overcome its binding energy, we can write the wind loss rate to be
\begin{equation}
\dot M_{{\rm wind}}=-f\frac {M_{{\rm NS}}R_{{\rm d}}^{3}\dot M_{{\rm NS}}}{4a^{2}R_{{\rm NS}}M_{{\rm d}}},
\end{equation}
where $R_{{\rm d}}$ is the radius of the donor star. The wind-driving efficiency $f$ was first estimated by \citet{1993ApJ...410..281T} to be between $10^{-3}$ and $10^{-1}$. To simulate the formation of compact black-hole binaries, \citet{2006MNRAS.366.1415} assumed a smaller value of $f$ in the range of $10^{-5} - 10^{-3}$. Here we adopt $f\sim 10^{-3} - 10^{-2}$.

The total mass loss rate of the donor star can be expressed as
\begin{equation}
\dot M_{{\rm d}}=\dot M_{{\rm tr}}+\dot M_{{\rm wind}}.
\end{equation}
Thus, the angular momentum loss rate via mass loss from the NS and the donor star is
\begin{equation}
\dot J_{{\rm ML}}= \frac {(\beta \dot M_{{\rm tr}} M_{{\rm d}}^{2} + \dot M_{{\rm wind}} M_{{\rm NS}}^{2})}{M^{2}}  a^{2}\omega,
\end{equation}
where $M=M_{{\rm NS}}+M_{{\rm d}}$ is the total mass of the system, and $\omega=2\pi/P_{\rm orb}$ is the angular velocity of the binary.

We then consider angular momentum loss due to magnetic braking by adopting the formula proposed by \citet{2006MNRAS.366.1415}. Assuming that the wind material escapes at the boundary of the magnetosphere, the rate of angular momentum loss via magnetic braking is given by
\begin{equation}
\dot J_{{\rm MB}} = \dot M_{{\rm wind}} \omega_{{\rm d}} r_{{\rm m}}^{2} = -B_{{\rm s}}R_{{\rm d}}^{13/4}\sqrt {-\dot M_{{\rm wind}}}(GM_{{\rm d}})^{-1/4}\omega,
\end{equation}
where $r_{{\rm m}}$ is the magnetospheric radius and $B_{{\rm s}}$ is the surface magnetic field of the donor star. When there is no X-ray excited wind we adopt the traditional magnetic braking law proposed by \citet{1981A&A...100L...7V} for low-mass stars with a convective envelope.

Now we present an analytical derivation of the orbital evolution. The orbital angular momentum of the system is
\begin{equation}
J = M_{{\rm NS}} M_{{\rm d}} (\frac{Ga}{M})^{1/2}.
\end{equation}
From the equation above we can get
\begin{equation}
\frac{\dot a}{a} = \frac{2\dot J}{J} - \frac{2\dot M_{{\rm d}}}{M_{{\rm d}}}+ \frac{\dot M}{M}-\frac{2\dot M_{{\rm NS}}}{M_{{\rm NS}}},
\end{equation}
with $\dot{x}=dx/dt$. Then we use a parameter $\alpha$ to represent the relation between $\dot M_{{\rm tr}}$ and $\dot M_{{\rm wind}}$, that is, $\dot M_{{\rm wind}} = \alpha \dot M_{{\rm tr}}$, and rewrite Eq.~(9) to be
\begin{equation}
\frac {\dot a}{a} = -\frac{2\dot M_{{\rm d}}}{M_{{\rm d}}}\left[1-\frac{(1-\beta)}{(1+\alpha)}q-\frac{(\alpha+\beta)M_{{\rm d}}}{2(1+\alpha)M}-\frac{\beta M_{{\rm d}}}{(1+\alpha)M}q- \frac {\alpha M_{{\rm NS}}}{(1+\alpha)M}\right]+\frac {2\dot J_{{\rm MB}}}{J}.
\end{equation}
A rough estimate shows that, if we take $M_{{\rm NS}} = 1.69\,M_{\sun}$, $M_{{\rm d}}= 0.46\,M_{\sun}$, and $\dot M_{{\rm tr}}=\dot M_{{\rm wind}} = 1\times 10^{-7}\,M_{\sun}$ yr$^{-1}$, then $\dot{P}_{\rm orb} \sim 10^{-10}$ s\,s$^{-1}$, which is of the same order of magnitude as the observed value.

\section{Numerical results} \label{sec:floats}

\subsection{The reference model}
We first construct our reference model with the following initial parameters. The NS mass is set to be $1.4\,M_{\sun}$, which actually has limited effect on the result. We start with a $1.1\,M_{\sun}$ donor star and a 0.4 d orbital period, and adopt a wind-driving efficiency of $5\times10^{-3}$. Recently, \citet{2017NatAs...1E.184S} showed that even M dwarfs and partly convective stars can generate strong magnetic fields of a few kilogauss, so we take the magnetic field of the donor star to be $900\,\rm G$.

\begin{figure}[ht!]
\plottwo{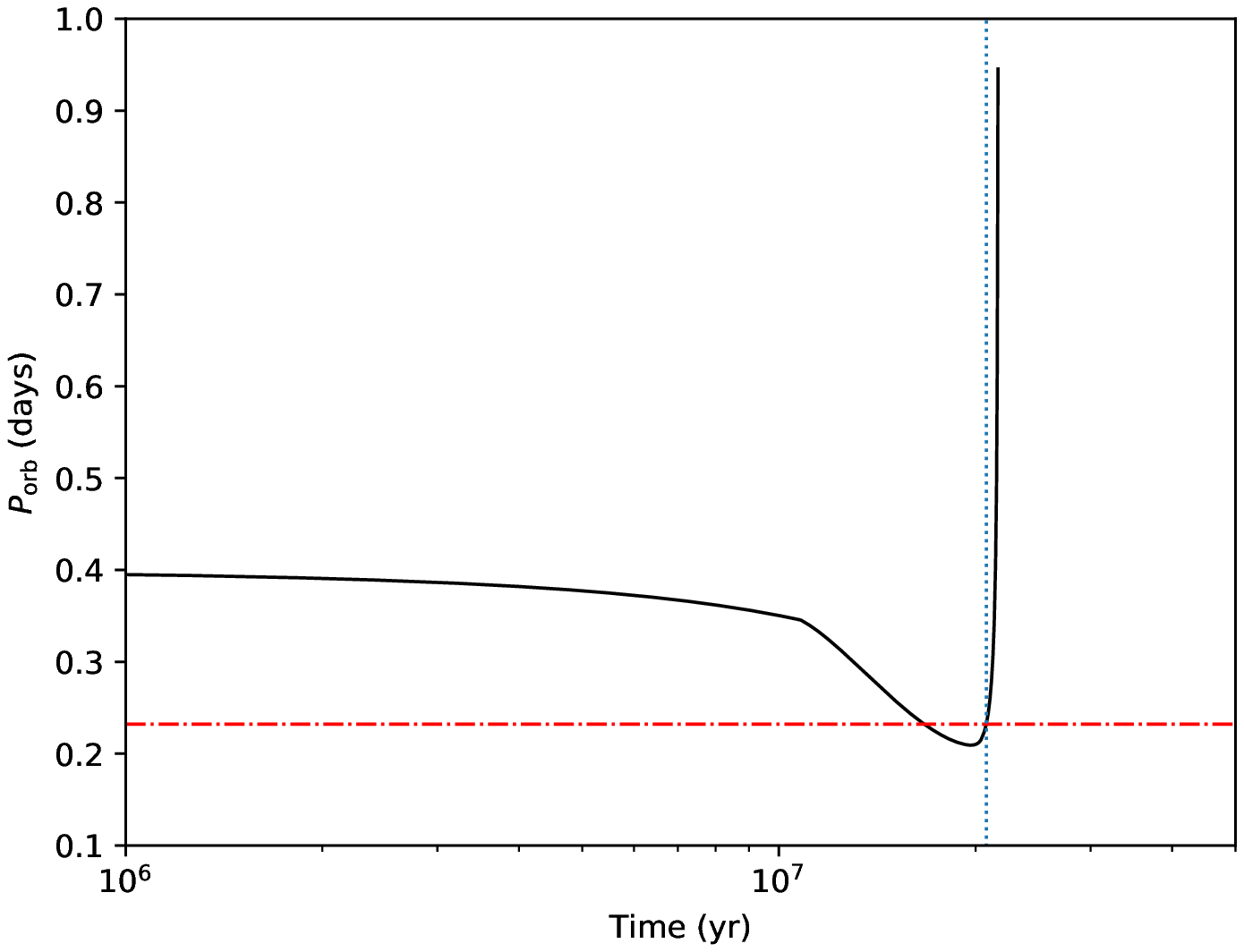}{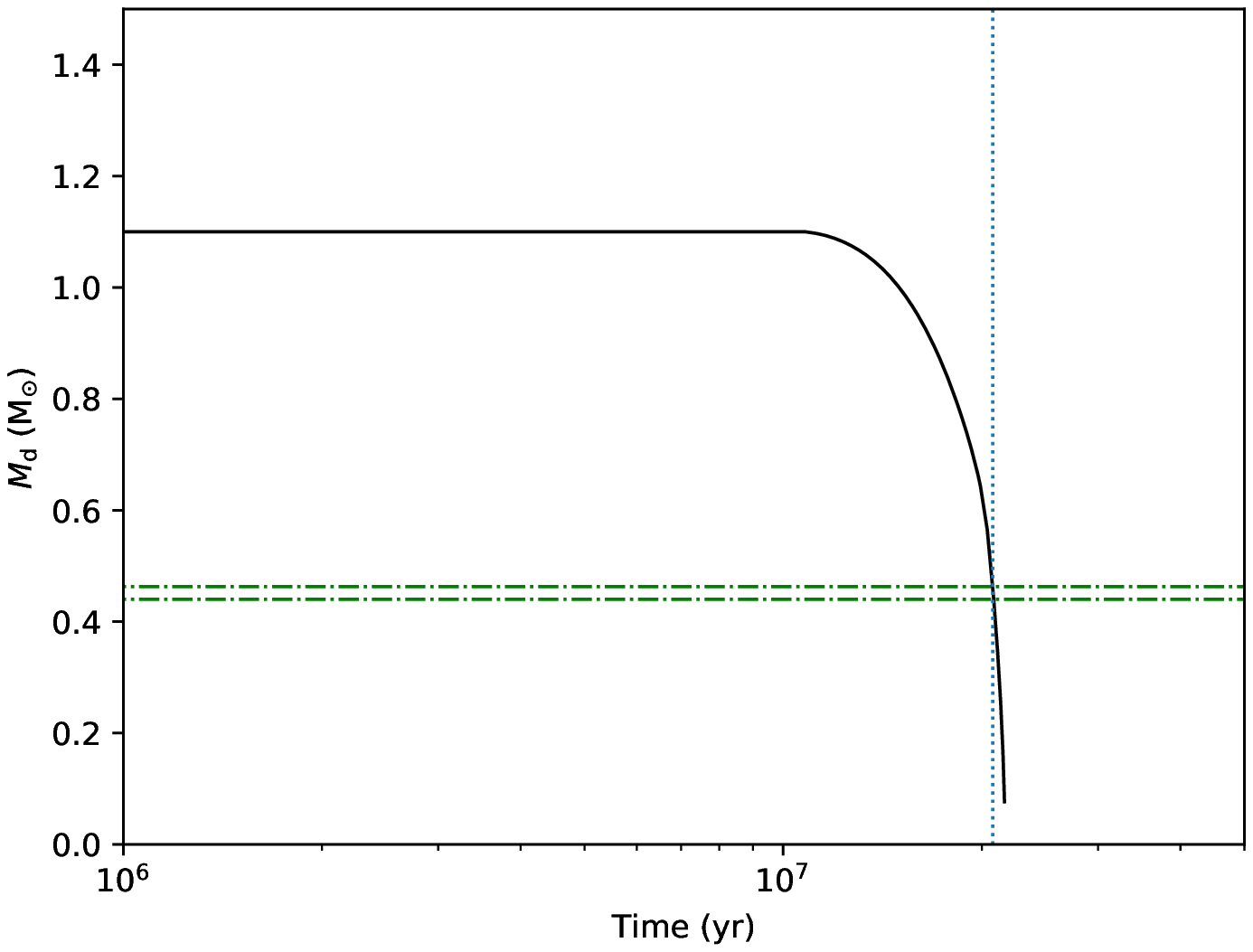}
\caption{Evolution of the orbital period (left panel) and the donor mass (right panel). The initial donor mass, orbital period, wind-driving efficiency, and surface magnetic field of the donor star are taken to be $1.1\,M_{\sun}$, $0.4\,\rm d$, $5\times 10^{-3}$, and $900\,\rm G$, respectively. The horizontal red dot-dashed line and green lines represent the measured value of the orbital period and the range of the  donor mass, respectively. The vertical dotted lines represent our calculated age of $2.096\times 10^{7}$ yr for 2A 1822$-$371.\label{fig:f1}}
\end{figure}

\begin{figure}
\plotone{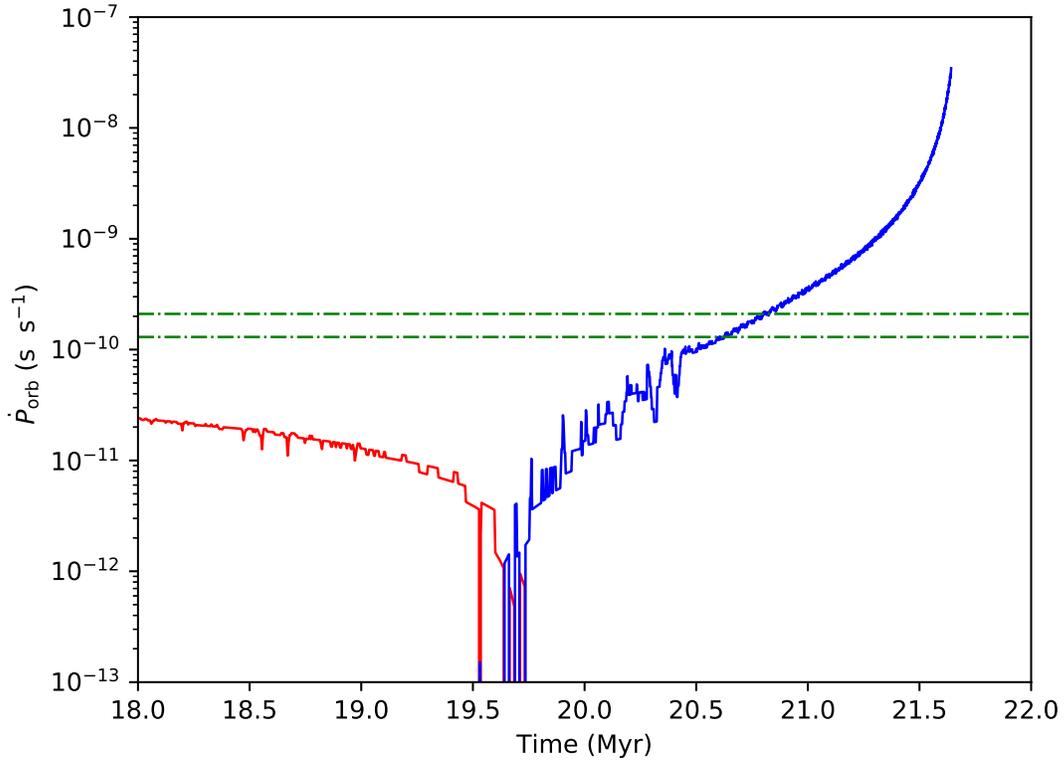}
\caption{The absolute value of the derivative of the orbital period as a function of time. The red line represents the negative values and the blue line positive values. The horizontal green lines confine the measured range of the period derivative. The initial parameters are same as in Fig.~\ref {fig:f1}.\label{fig:f2}}
\end{figure}

The evolution of the orbital period and the donor mass is shown in the left and right panels of Fig.~\ref {fig:f1}, respectively. The horizontal dashed lines represent the measured value of the orbital period and the range of the donor mass. They cross the evolutionary paths at the age of $2.096\times 10^{7}$ yr, indicating that 2A 1822$-$371 is a relatively young system. The orbital evolution is initially driven by angular momentum loss via mass loss and magnetic braking with decreasing orbital period. At the age of $\sim 2\times 10^7$ yr the orbit starts to expand because the mass transfer rate becomes sufficiently high.
The evolution of the orbital period derivative is shown in Fig.~\ref {fig:f2}, with the red and blue lines denoting negative and positive values, respectively. The two horizontal lines represent the range of the measured orbital period derivative.

\begin{figure}
\plotone{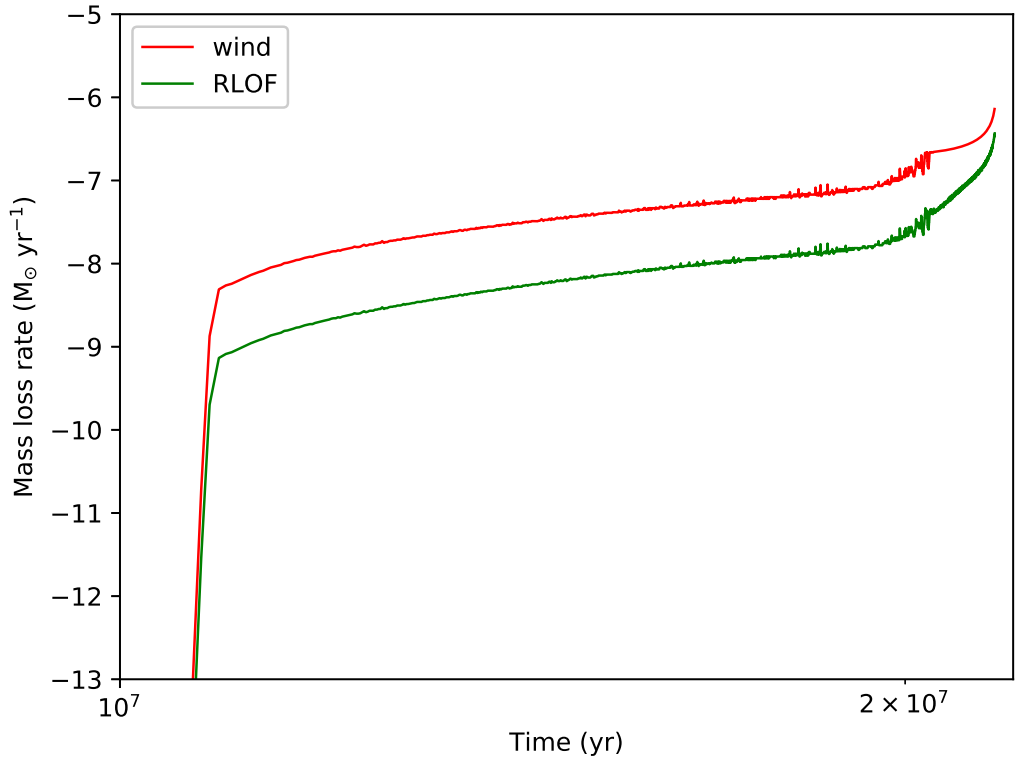}
\caption{Evolution of the RLOF mass-transfer rate (green line) and the wind-loss rate (red line). The initial parameters are same as in Fig.~\ref {fig:f1}. \label{fig:f3}}
\end{figure}

The total mass loss rate of the donor star is composed of the excited wind loss rate and the mass transfer rate through RLOF. They are depicted with the red and green lines in Fig.~\ref{fig:f3}, respectively. Accompanied by decreasing donor mass, the binding energy of the donor star keeps decreasing, leading to a stronger wind excited by X-ray irradiation. The wind-driven mass transfer rate increases with time and reaches $\sim 2\times 10^{-8}\,M_{\sun}$ yr$^{-1}$ when $M_{\rm d}\simeq 0.46\,M_{\sun}$. In the final evolutionary stage the mass transfer becomes unstable when the donor evolves to be a brown dwarf if X-ray irradiation keeps working.

\subsection{Parameter study}
We have shown that it is possible to reproduce the properties (orbital period and its derivative, donor mass and mass transfer rate) of 2A 1822$-$371 with the parameters in the reference model. In this subsection we study the effect of the initial parameters on the binary evolution. To achieve this, we change one parameter once and keep other parameters same as in the reference model.

\begin{figure}
\plotone{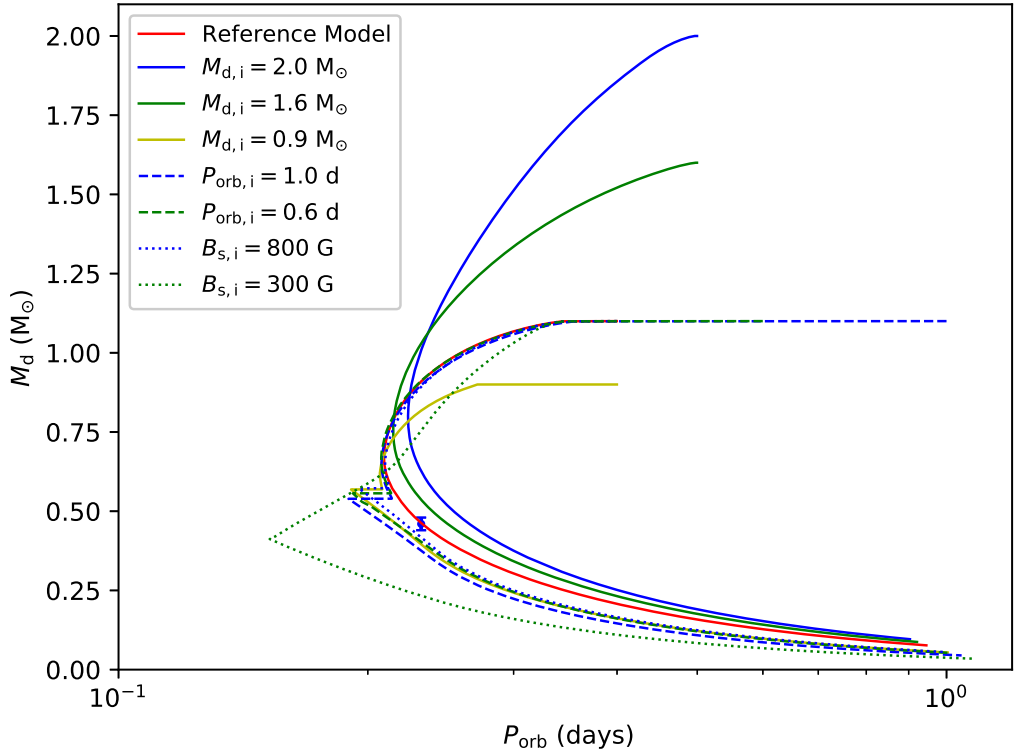}
\caption{Evolutionary tracks in the donor mass - orbital period plane with varied initial parameters. The solid red line represents the reference model. Other solid lines show the cases with different initial donor masses, the dashed and dotted lines show the cases with different initial orbital periods and magnetic field of donors, respectively.  The dot indicates the observed values of 2A 1822$-$371.\label{fig:f4}}
\end{figure}

We first consider the influence of the initial donor mass by adopting $M_{\rm d}=0.9\,M_{\sun}$, $1.1\,M_{\sun}$, $1.6\,M_{\sun}$, and $2.0\,M_{\sun}$ for comparison. Fig.~\ref {fig:f4} shows the orbital evolution with the donor mass for varied initial parameters. Note that for donor stars with $M_{\rm d}=1.6\,M{_\sun}$ and $2.0\,M_{\sun}$ we set the initial period to be $0.5\,\rm d$, because the companion star overflows its RL at the beginning with a 0.4 d orbital period. We can see that in all cases with different initial donor mass, the general evolutionary trends are similar, with more massive initial donor star leading to heavier companion star at the same orbital period.

Fig.~\ref{fig:f4} also compares the evolutionary tracks in the donor mass - orbital period plane with the initial orbital period being 0.4 d, 0.6 d, and 1.0 d. In the latter two cases, the mass transfer suspends for some time, leading to a deviation from the evolutionary track with $P_{\rm orb}=0.4$ d. Then the evolutionary tracks coincide at the later stage. Thus, a slight change of the initial period has limited influence on the results.

The efficiency of magnetic braking is sensitively dependent on the donor's magnetic field. We vary $B_{\rm s}$ from $900\,\rm G$ to $800\,\rm G$ and $300\,\rm G$ to investigate the influence of the donor's magnetic field.  When $B_{\rm s}$ becomes weaker, angular momentum loss due to the magnetized wind decreases, and so does the mass loss rate. The orbital-period expansion rate becomes smaller correspondingly. This results in a smaller orbital period at the same donor mass during the orbital expansion phase, as depicted in Fig.~\ref{fig:f4}.

\begin{figure}
\plottwo{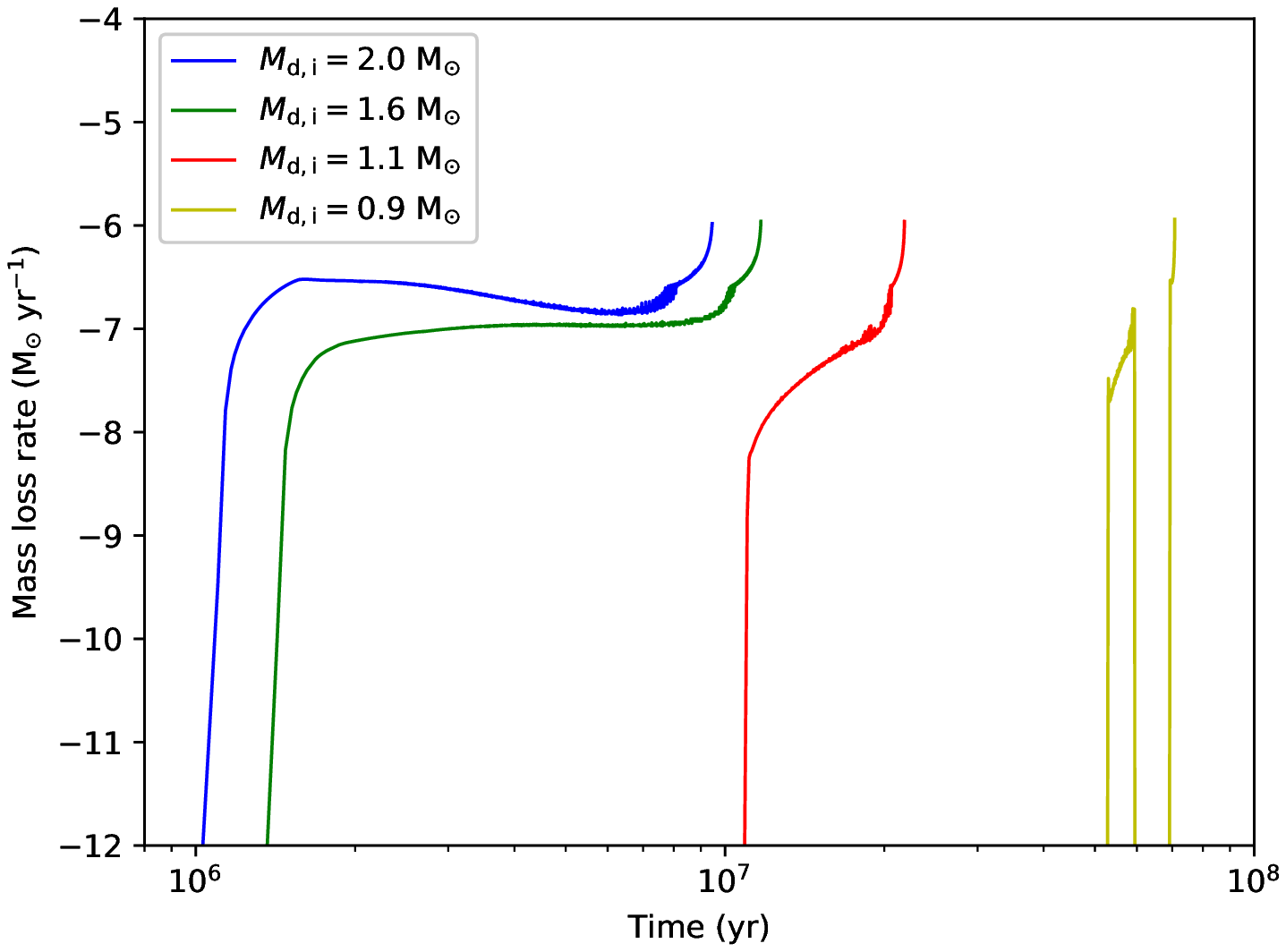}{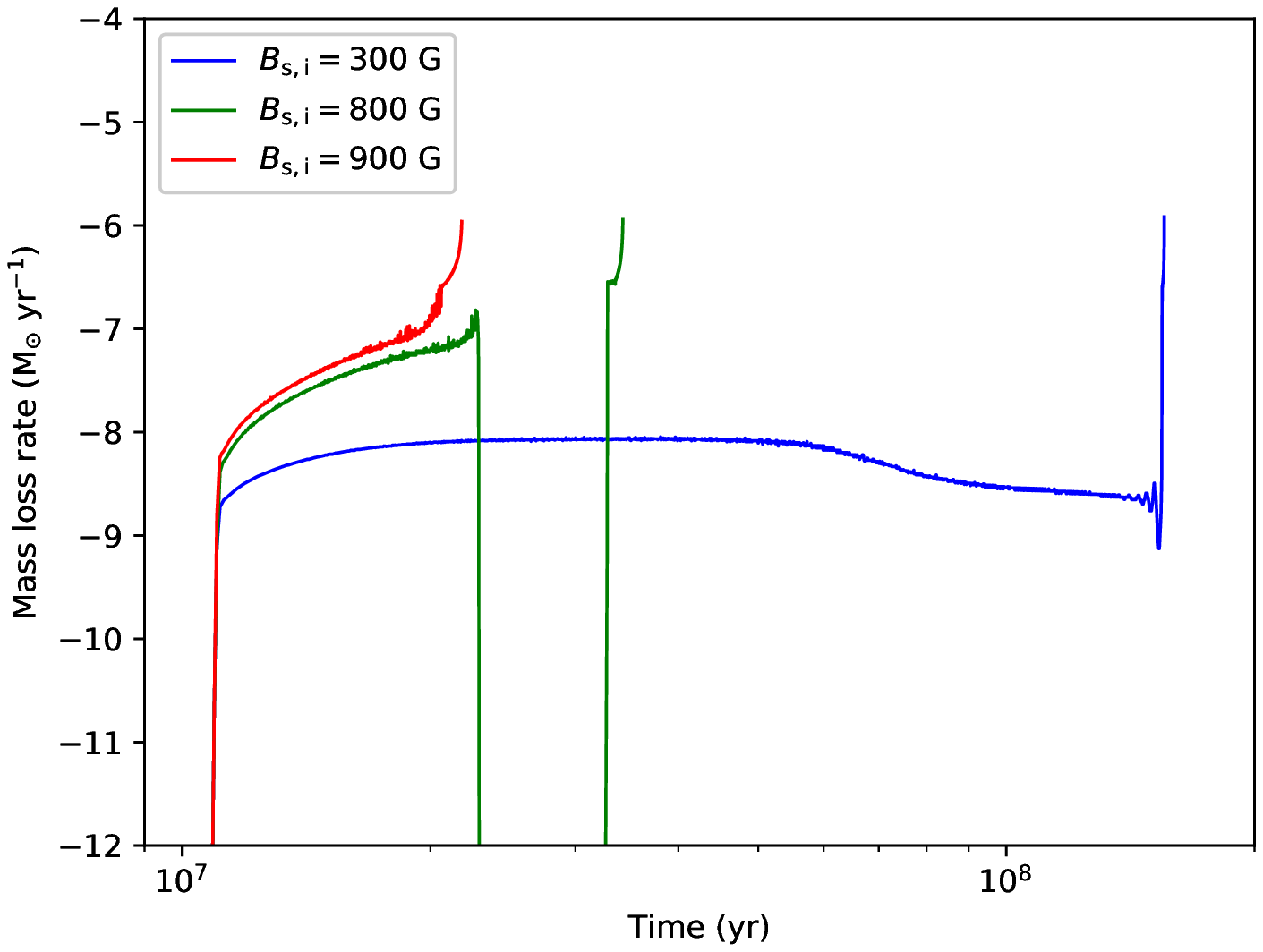}
\caption{Left panel: evolution of the mass loss rate with different initial donor mass. The yellow, red, green, and blue curves represent the initial donor masses of 0.9, 1.1, 1.6 and $2.0\,M_{\sun}$, respectively. The wind-driving efficiency $f$ =  $5\times 10^{-3}$ and the initial magnetic field $B_{{\rm s}} = 900\,\rm G$. The initial orbital period is $0.4\,\rm d$ for the cases with the initial donor masses of $0.9\,M_{\sun}$ and $1.1\,M_{\sun}$, and $0.5\,\rm d$ for the cases with the initial donor masses of $1.6\,M_{\sun}$ and $2.0\,M_{\sun}$. Right panel: evolution of the mass loss rate with initial donor mass of $1.1\,M_{\sun}$ but with different initial magnetic field of the donor. The blue, green, and red curves represent the initial magnetic field of 300, 800, and $900\,\rm G$, respectively. The initial orbital period is $0.4\,\rm d$, and the wind-driving efficiency $f$ =  $5\times 10^{-3}$.\label{fig:f5} }
\end{figure}

The left panel of Fig.~\ref{fig:f5} demonstrates the mass loss rate of the system with different initial donor mass. We can see that the evolutionary trends are similar except that in the case of $0.9\,M_{\sun}$ the system undergoes a temporary detached phase for $\sim 10^{7}$ yr. It occurs at the time when the orbit starts to expand. And we plot the corresponding mass loss rate with varied donor's magnetic field in the right panel of Fig.~\ref{fig:f5}. When $B_{\rm s}$ becomes weaker, angular momentum loss due to the magnetized wind decreases, and so does the mass loss rate.

\begin{figure}
\plotone{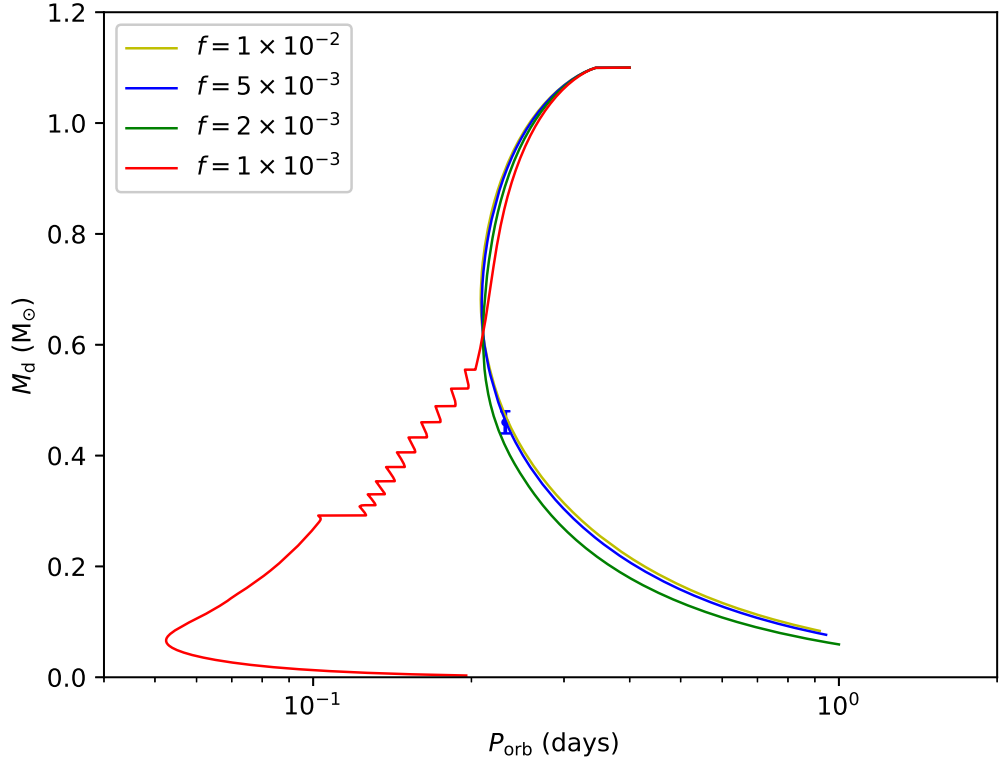}
\caption{Same as Fig.~4 but for different wind-driving efficiencies. The yellow, blue, green and red curves represent the wind-driving efficiency of $1\times 10^{-2}$, $5\times 10^{-3}$, $2\times 10^{-3}$, and $1\times 10^{-3}$, respectively. The initial donor mass $M_{{\rm d}}=1.1\,M_{\sun}$, the initial magnetic field $B_{{\rm s, i}} = 900\,\rm G$, and the initial orbital period $P_{{\rm orb, i}} =0.4\,\rm d$. \label{fig:f6}}
\end{figure}

We finally discuss the effect of the wind-driving efficiency $f$. Fig.~\ref{fig:f6} shows the evolutionary tracks with $f=1\times 10^{-3}$, $2\times 10^{-3}$, $5\times 10^{-3}$, and $1\times 10^{-2}$. Except for the first case, the binary evolutions in other cases seem able to reproduce the properties of 2A 1822$-$371. When $f=1\times 10^{-3}$, the orbital expansion phase starts much later when $M_{\rm d}\leq 0.1\,M_{\sun}$.

We need to point out that the effect of magnetic braking depends on  $f^{1/2}B_{\rm s}$, which means that $f$ and $B_{\rm s}$ work in a combined way. For example, if taking $f=1\times10^{-3}$  and $B_{\rm s}=1800\,\rm G$, we get similar results as in the reference model. except that the mass transfer rate and  the orbital period derivative are both a little lower. So there is degeneracy in constraining $f$ and $B_{\rm s}$ from the observed values.

\subsection{What happens if $B_{{\rm s}}$ evolves with $P_{orb}$?}
The structure of the donor star changes when it loses its material. The dynamo activity in the star may also change accordingly. Therefore, it is interesting to see what happens if $B_{{\rm s}}$ evolves with time. It is commonly assumed that the magnetic field strength evolves with the Rossby number, following the relation $B_{{\rm s}}\propto R_{{\rm o}}^{-\alpha}$, where the Rossby number $R_{{\rm o}}=2\pi/(\Omega\tau_{{\rm c}})$ and $\tau_{{\rm c}}$ is the convective turnover timescale. Observations of young Solar-type stars in open clusters and stellar associations of known ages suggested $\alpha=1.0\pm 0.1$ \citep{2016MNRAS.457..580F}. \citet{2017MNRAS.472.2590S} modeled the rotational behavior of Solar-type stars from the pre-main-sequence through the end of their main-sequence, and estimated $\alpha = 1.2$. In this work we adopt $\alpha = 1$ and the following relation \citep{2006ApJ...653L.137I} to determine the surface magnetic field of the companion star,
\begin{equation}
\frac {B_{{\rm s}}}{B_{{\rm s, i}}} = f_{{\rm b}}\frac{\tau_{{\rm c}}}{800\ \text{days}}\frac{1\ \text{day}}{P_{{\rm orb}}},
\end{equation}
where $B_{{\rm s, i}}$ is the initial magnetic field of the companion star and  $f_{{\rm b}}$ is a scaling factor. To evaluate $\tau_{{\rm c}}$, we use the empirical relation by \citet{2017MNRAS.472.2590S}:
\begin{equation}
\text{log}(\tau_{{\rm c}}) = 8.79 - 2|{\text{log}(m_{{\rm cz}})}|^{0.349}-0.0194|{\text{log}(m_{{\rm cz}})}|^{2}-1.62\text{min}[\text{log}(m_{{\rm cz}})+8.55,0] ,
\end{equation}
where $m_{{\rm cz}}$ represents the mass of the convective zone divided by the total stellar mass.

\begin{figure}
\plotone{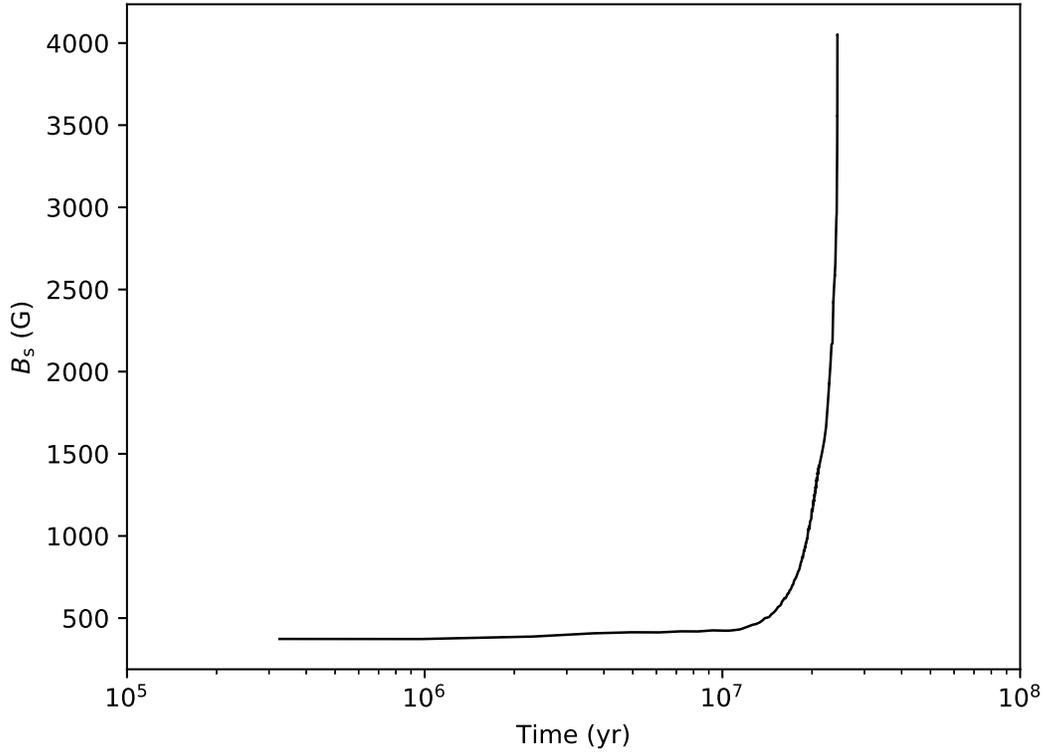}
\caption{Evolution of the donor's surface magnetic field. Its initial value is taken to be $370\,\rm G$.\label{fig:f7}}
\end{figure}

\begin{figure}[ht!]
\plottwo{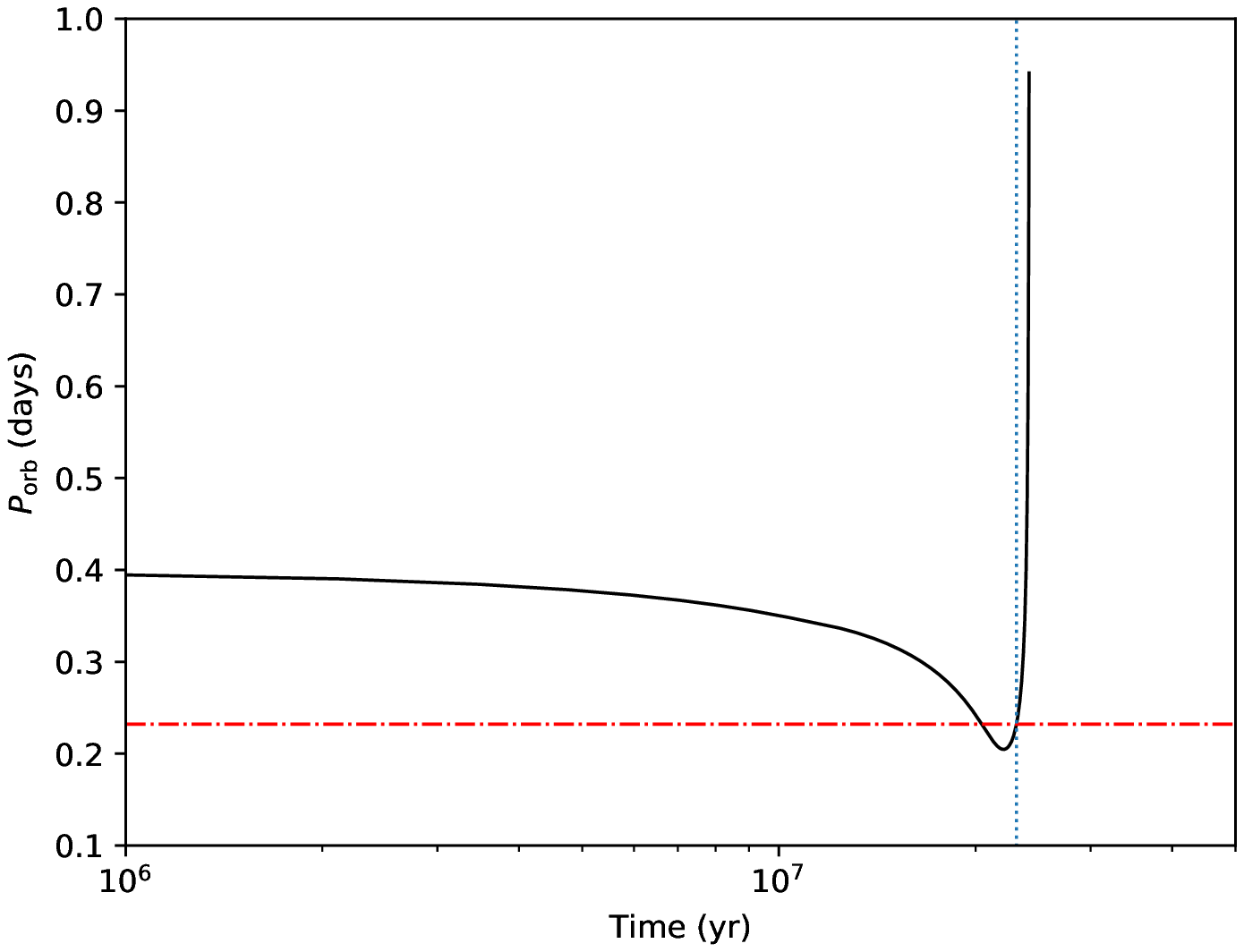}{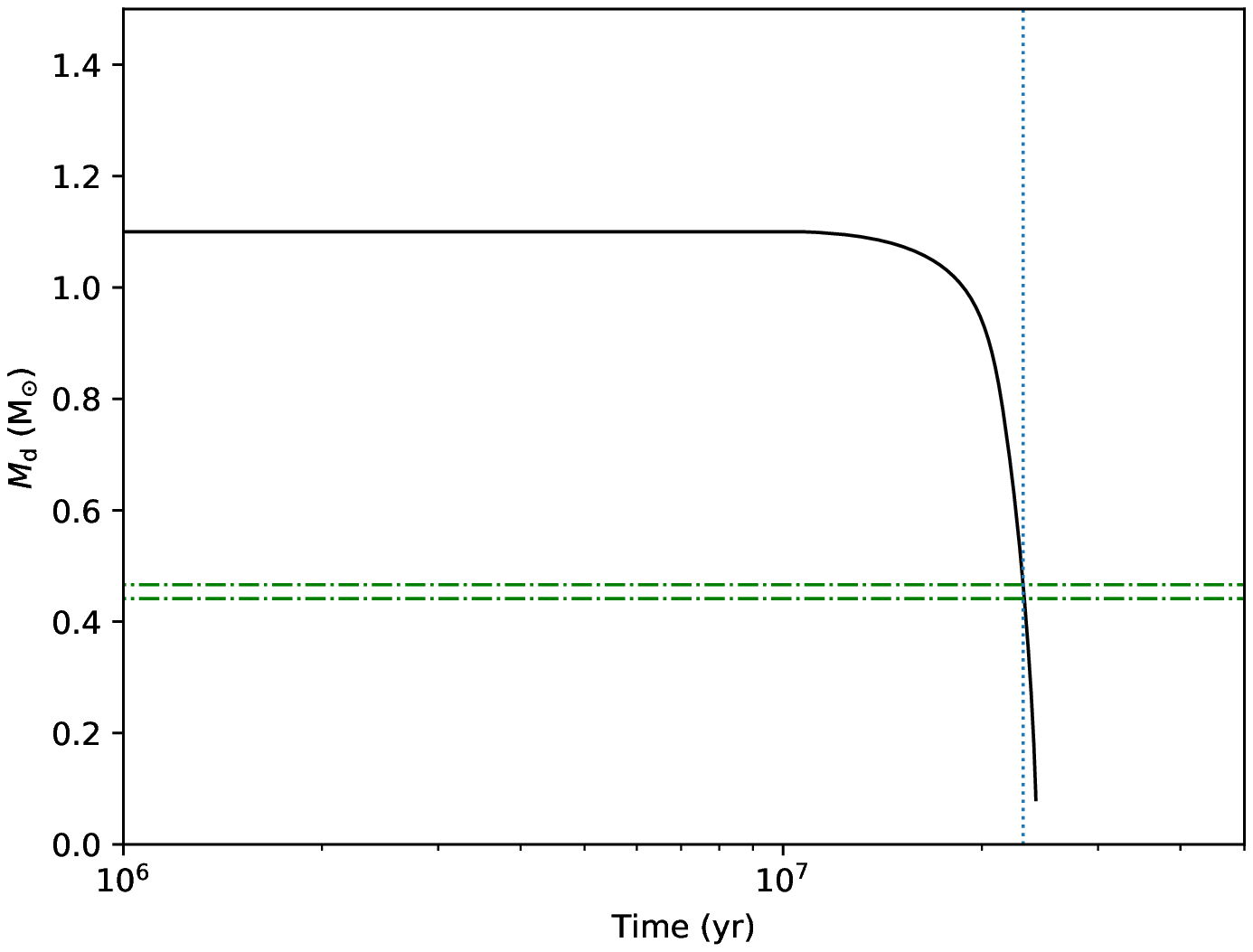}
\caption{Evolution of the orbital period (left panel) and the donor mass (right panel) when the magnetic field evolution is taken into account. The initial magnetic field $B_{{\rm s, i}}=370\,\rm G$, and other parameters are the same as in  Fig.~\ref {fig:f1}. The horizontal red dot-dashed lines represent the observed value of 2A 1822$-$371 and the green horizontal lines represent the range of the donor mass. The vertical dotted lines represent our calculated age of $2.233\times 10^{7}$ yr for 2A 1822$-$371. \label{fig:f8}}
\end{figure}

\begin{figure}
\plotone{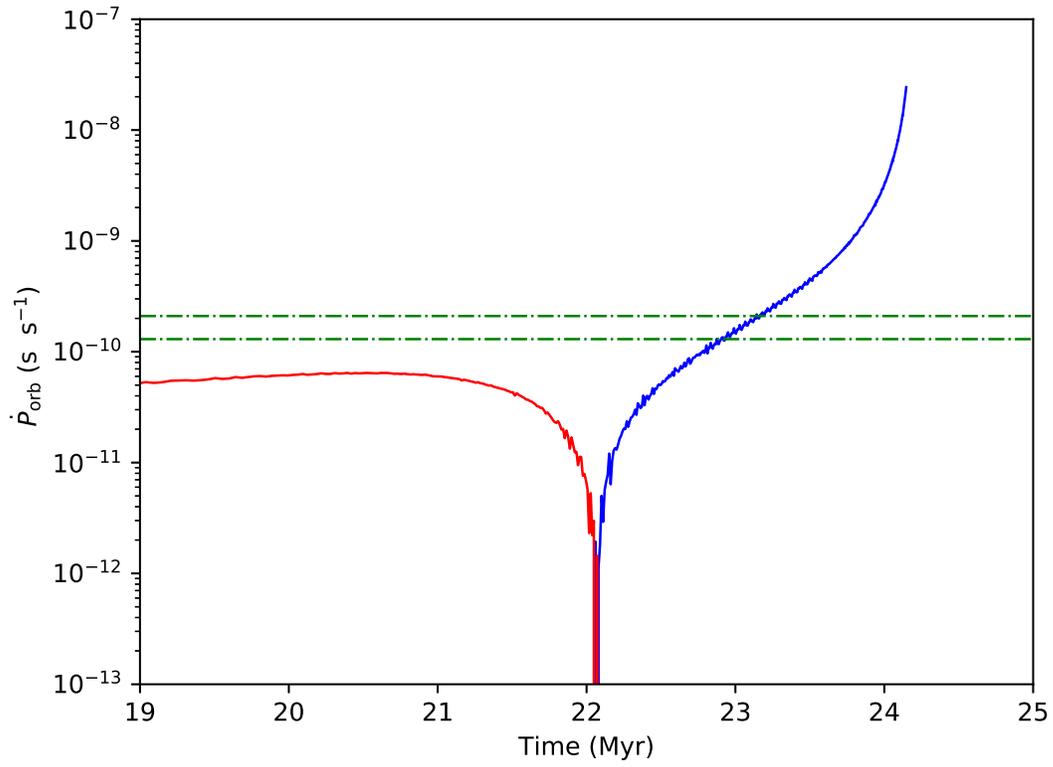}
\caption{Same as Fig.~2 but for an evolving magnetic field. The initial parameters are the same as in Fig.~\ref {fig:f8}.\label{fig:f9}}
\end{figure}

In our model, the donor star loses mass efficiently, leading to a noticeable increase in $\tau_{{\rm c}}$. Although $P_{{\rm rot}}$ grows with the expanding orbit, $B_{{\rm s}}$ keeps increasing during the evolution. As a result, it requires a lower initial $B_{{\rm s, i}}$ as well as a smaller $f$ to fit the observations. We display an example with $f=3\times10^{-3}$ and $B_{{\rm s, i}}= 370\,\rm G$. The evolution of the magnetic field is shown in Fig.~\ref {fig:f7}, and we can see that it increases with time as expected.  Fig.~\ref {fig:f8} shows the evolution of donor mass and orbital period with time, and the evolution of the derivative of the orbital period is plotted in Fig.~\ref{fig:f9}. At $2.233\times 10^{7}$ yr, the orbital period of the system is 0.232 d, the donor-mass is $0.466\, M_{\sun}$ and the orbital period derivative is $1.962\times 10^{-10}$ s\,s$^{-1}$.


\section{Discussion} \label{sec:displaymath}
The high mass transfer rate and rapidly expanding orbit of 2A 1822$-$371 have been a mystery, which are inconsistent with the traditional LMXB evolution model. More generally, it has been shown that the inferred mass transfer rates of some short-period LMXBs are at least an order of magnitude higher than theoretically expected mass transfer rates for the same orbital periods \citep{2002ApJ...565.1107P}. This discrepancy has stimulated new ideas on the modification of the efficiency of magnetic braking in LMXBs. For example, by applying a magnetic braking prescription which accounts for increased wind mass-loss in evolved stars compared to main sequence stars, \citet{2016MNRAS.456..263P} and more recently \citet{van2019} suggested some modified magnetic braking laws to account for the observed accretion rates. However, they still face difficulties in  simultaneously reproducing the accretion rate, mass ratio, and rate of orbital expansion for 2A 1822$-$371. We have shown that  both rapid mass transfer and orbital expansion in 2A 1822$-$371 requires some extra angular momentum loss, such as magnetic braking due to irradiation-driven mass loss. This does not mean that the evolution of 2A 1822$-$371 should follow the scenario described in this work, because there are still some issues to be addressed. For example, it requires that the donor star possess a strong magnetic field, and the efficiency of irradiation-driven wind mass loss is quite uncertain. However, the irradiation-driven mass loss model presents an alternative evolutionary channel for some LMXBs,  and it is interesting to discuss its possible implications.

In the traditional model, the orbital evolution of LMXBs mainly depends on the competition between angular momentum loss due to magnetic braking and/or gravitational radiation and angular momentum transfer caused by mass transfer. The former tends to shrink the orbit and the latter to expand it. So there is a so-called bifurcation period $P_{{\rm bif}}$, which can be defined as the initial orbital period separating the converging and diverging systems.
\citet{1988A&A...191...57P,1989A&A...208...52P} for the first time did systematic investigations on the bifurcation period. Neglecting mass loss from the binary system and using the \citet{1981A&A...100L...7V} magnetic braking law, they found that the bifurcation period is around $0.5-1.0$ d. \citet{1998MNRAS.300..352E} recalculated the bifurcation period by considering mass loss from the binary system. \citet{2002ApJ...565.1107P} defined the bifurcation period as the orbital period when RLOF begins and found that it is around 18 hr for a $1.4\,M_{\sun}$ NS with a $1\,M_{\sun}$ companion star. \citet{2005A&A...440..973V,2005A&A...431..647V} showed that the bifurcation period strongly depends on the magnetic braking efficiency, and \citet{2009ApJ...691.1611M} calculated the bifurcation periods to be $10-30$ hr for binary systems with a $1.4\,M_{\sun}$ NS and a $0.5-2\,M_{\sun}$ donor star, taking into account various kinds of magnetic braking and mass-loss
mechanisms.

\begin{figure}
\plotone{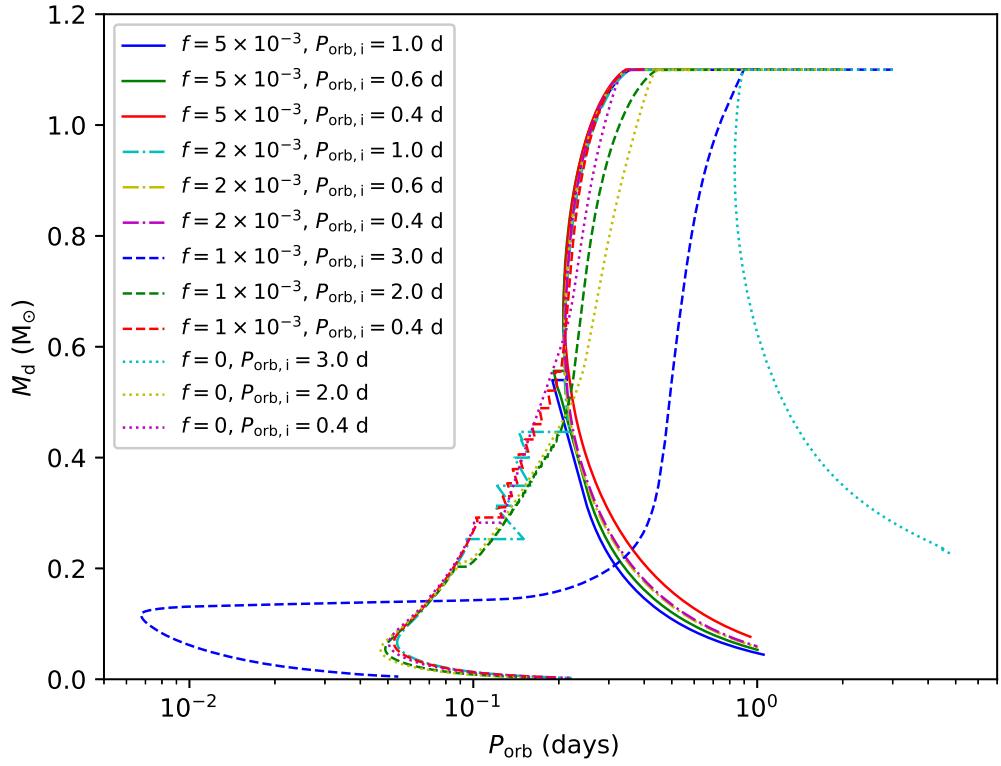}
\caption{Same as Fig.~4 but for different initial orbital periods and wind-driving efficiencies. The initial donor mass $M_{{\rm d}}=1.1\,M_{\sun}$ and the initial magnetic field $B_{{\rm s}} = 900\,\rm G$.\label{fig:f10}}
\end{figure}

For the LMXB evolution with irradiation-driven mass loss, the bifurcation period could be significantly changed. Fig.~\ref {fig:f10} compares the $M_{{ \rm d}}-P_{{\rm orb}}$ relations for different values of the wind efficiency $f$. The solid, dot-dashed, dashed, and dotted lines correspond to $f=5\times10^{-3}$,  $2\times10^{-3}$, $1\times10^{-3}$, and 0, respectively. The initial orbital period $P_{\rm orb,i}$ varies from $0.4\,\rm d$ to $3.0\, \rm d$. We fix the initial donor mass $M_{\rm d,i} = 1.1\,M_{\sun}$ and the initial magnetic field $B_{\rm s} = 900\,\rm G$.
In the case of $f=0$, the orbital evolution clearly shows diverse paths with different initial orbital periods, that is, initially long-period binary becomes wider and short-period binary becomes more compact. The situation changes when $f=1\times10^{-3}$. For $P_{\rm orb,i}=3.0\,\rm d$, the binary orbit eventually shrinks rather expands because angular momentum loss due to irradiation-driven wind is much more efficient than the traditional magnetic braking effect. When $f$ is increased to $2\times10^{-3}$ and $5\times10^{-3}$, the evolution changes drastically. Even for $P_{\rm orb,i}=0.4\,\rm d$, the orbital period evolves to be nearly $1.0\,\rm d$ at the late stage. The reason is that mass transfer at this stage tries to widen the orbit, and largely compensates the shrinking tendency caused by the wind.  This means that the traditional bifurcation period does not apply in these cases. Because of the high mass transfer rates the binaries are stable against the thermal-viscous instability in the accretion disks, so they should appear as persistent sources. This is compatible with the fact that most persistent NS LMXBs have main-sequence donors and orbital periods between $1-100$ hr \citep{R11,C12}. Finally, the NSs will turn into millisecond pulsars when the mass transfer ceases. Their orbital periods may cover several hours to around 1 day, which are difficult to attain in the traditional model for LMXB evolution \citep{SL15}. Traditional treatment of magnetic braking predicts a paucity of binary millisecond pulsars (BMPs) with orbital periods around one day. However, this is not seen in the real distribution \citep[also][]{2003ApJ...597.1036P,2011ApJ...732...70L,2014ApJ...791..127J}. \citet{2014A&A...571A..45I} summarized several effects may potentially affect the LMXB evolution, such as irradiation of the donor star, accretion disk instabilities and a circumbinary disk. Our results suggest that irradiation-driven wind could significantly influence the angular momentum transfer in LMXBs, helping solve this problem.

Another possible mechanism related to irradiation is the mass transfer cycles in LMXBs. The X-ray irradiation incident onto the companion star may drive an increase in the internal energy of the companion. The companion star would expand on thermal time-scale when absorbing the radiation energy, resulting in a rapid mass transfer. This process could lead to an orbital expansion, and, after rapidly transferring a mount of matter, the binary becomes nearly detached for a long period \citep{2004A&A...423..281B}. \citet{2003ApJ...597.1036P} suggest that irradiation-induced mass-transfer cycles can reduce the X-ray active lifetime while increasing the luminosity. This may account for the discrepancy between the theoretical and measured accretion rates of LMXBs \citep{2002ApJ...565.1107P} and the discrepancy between the birthrates of LMXBs and BMPs \citep{1988ApJ...335..755K}.
Note also that during the low state the NS may act as a dim X-ray source. Moreover, since the mass transfer rate is very low, the accretion disk may be subject to thermal instability and the accreting NS becomes transient, with the peak luminosities during outbursts less than $\sim 10^{36}$ ergs$^{-1}$, like so-called very faint X-ray transients  \citep[VFXTs,][]{w2006}. Alternatively, accretion may be completely prohibited by the radiation pressure of the NS, which then appears as a millisecond pulsar. It may evaporate its companion with high-energy radiation \citep{1989ApJ...343..292R}, so the binary may evolve to be a redback \citep{R13}. If this picture applies for 2A 1822$-$371-like objects, we expect that either VFXTs or redbacks should be much more abundant than luminous LMXBs with similar orbital periods and companion masses, since the (semi-)detached stage is much longer than the rapid mass transfer stage. Unfortunately, the currently known numbers of VFXTs and redbacks prevent a solid statistical inference.

\section{Summary}
We investigate the evolution of 2A 1822$-$371 employing an excited wind-driving mass-transfer model. Assuming a strong magnetic field of the donor and a high wind-driving efficiency, we find that mass transfer can proceed rapidly, thus simultaneously account for the donor mass, fast orbital expansion and high X-ray luminosity of 2A 1822$-$371. Our numerically simulated results show that the total mass loss rate of the donor star is $\sim 10^{-7}\,M_{\sun}$ yr$^{-1}$ when the donor mass is around $0.5\,M_\odot$, consistent with the measured value.

Depending on the wind-driving efficiency, the binary evolution can change significantly. On one hand the wind mass loss strengthens the effect of magnetic braking; on the other hand, the induced mass transfer tries to widen the orbit. Hence in some situations the balance between the two factors leaves the final orbital period close to its original value. This scenario provides a possible way to explain the formation of millisecond pulsars with orbital periods between several hours and about one day.


\acknowledgments
This work was supported by the National Key Research and Development Program of China (2016YFA0400803), the Natural Science Foundation of China under grant No. 11773015 and Project U1838201 supported by NSFC and CAS.

%

\vspace{5mm}

\end{document}